\documentclass[10pt,final,journal,twocolumn]{IEEEtran}

\ifCLASSINFOpdf
\else
\fi

\usepackage{cite}
\usepackage[cmex10]{amsmath}
\usepackage{amsmath}
\usepackage{algorithm}
\usepackage{algorithmic}
\usepackage{stfloats}
\usepackage{multicol,multienum}
\usepackage{multirow}
\usepackage[dvips]{graphicx}
\usepackage[caption=false,font=footnotesize]{subfig}
\usepackage{wrapfig}
\usepackage{color}

\begin{document}
\title{Anti-jamming Communications Using Spectrum Waterfall: A Deep Reinforcement Learning Approach}

\author{
       Xin~Liu,
       Yuhua~Xu,~\IEEEmembership{Member,~IEEE},
       Luliang~Jia,~\IEEEmembership{Student Member,~IEEE},
       Qihui~Wu,~\IEEEmembership{Senior Member,~IEEE,}
       and Alagan~Anpalagan,~\IEEEmembership{Senior Member,~IEEE}




\thanks{
X.~Liu is with the College of Information Science and Engineering, Guilin University of Technology, Guilin 541004, China. (e-mail:leo$\_$nanjing@126.com).}

\thanks{
 Y.~Xu and L.~Jia are with the College of Communication Engineering, PLA
Army Engineering University, Nanjing 210007, China. (e-mail:
 yuhuaenator@gmail.com;jiallts@163.com).}

\thanks{Qihui Wu is  with the College of Electronic and Information Engineering,
Nanjing University of Aeronautics and Astronautics, Nanjing, China (e-mail: wuqihui2014@sina.com).}

\thanks{Alagan~Anpalagan is with the Department of Electrical and Computer Engineering, Ryerson University, Toronto, Canada (alagan@ee.ryerson.ca).}
 }

\maketitle

\begin{abstract}
This letter investigates the problem of anti-jamming communications in dynamic and unknown environment through on-line learning. Different from existing studies which need to know (estimate) the jamming patterns and parameters, we use the spectrum waterfall, i.e., the raw spectrum environment, directly. Firstly, to cope with the challenge of infinite state of raw spectrum information, a deep anti-jamming Q-network  is constructed. Then, a  deep anti-jamming reinforcement learning  algorithm is proposed to obtain the optimal anti-jamming strategies.  Finally, simulation results validate the the proposed approach. The proposed approach is relying only on the local observed information and does not need to estimate the jamming patterns and parameters, which implies that it can be widely used  various anti-jamming scenarios.

\end{abstract}


\IEEEpeerreviewmaketitle

\vspace{-0.2cm}
\section{Introduction}
Anti-jamming is always an active research topic, as wireless transmissions are naturally vulnerable to jamming attacks. The mainstream anti-jamming  techniques  includes Frequency Hopping Spread Spectrum (FHSS) and Direct-Sequence Spread Spectrum (DSSS) \cite{United_Against}. Recently, to address the interactions between the legitimate users and the jammers, game theory has been widely applied in the literature \cite{bb_gamelearning1, bb_gamelearning2, bb_gamelearning3, bb_sbgame2, bb_bysbgame1, bb_bysbgame2}. In methodology,  these approaches need to know the jamming strategies, which implies that the legitimate users are required to estimate the jamming patterns and parameters from the observed environment. However,  with the rapid development of artificial intelligence and universal software radio peripheral (USRP) \cite{You_Can_Jam}, the jammers can easily create dynamic and intelligent jamming attacks. As a consequence, there are two limitations with regard to estimation-based anti-jamming communications: i) there may be information loss for unknown jamming patterns, and ii) if the intelligent jammer switches its strategies  dynamically and rapidly, it is not possible to track and react it in real time. Thus, it is challenging and interesting to investigate anti-jamming communication approaches in dynamic and unknown environment.

To overcome the above limitations, a promising  way is to design new anti-jamming approaches that utilize the raw environmental information, which is known as spectrum waterfall \cite{bb_waterfall}, without estimating jamming patterns and parameters. These kind of anti-jamming approaches would avoid information loss and adapt to the dynamic environment, as can be expected. In addition,  online learning is an effective way to solve the decision problems in dynamic environment.  The widely used technique is  Q-learning \cite{bb_Qlearning}, which has been used in anti-jamming problems \cite{bb_gamelearning1, bb_gamelearning2}. Unfortunately, Q-learning is not able to deal with the raw environmental information directly because of the infinite state of the environment.

Motivated by the deep reinforcement learning technique for learning successful control policies from raw video data in \cite{bb_DQN}, we investigate the anti-jamming problem in unknown and dynamic environment. First, the raw spectrum information is defined as the state of the environment to avoid losing the jammer information as much as possible.
 Then, a deep anti-jamming Q-network (DAQN) is constructed to realize the direct processing of raw spectrum information. Finally, a deep anti-jamming reinforcement learning algorithm (DARLA) is proposed.  Simulation results show that the proposed DARAL achieves the best anti-jamming strategies in various scenarios. The main contributions are summarized as follows.
\begin{itemize}
\item Based on the deep reinforcement learning technique, a smart anti-jamming communication scheme  is proposed.  In particular, the raw spectrum information is defined as a state, which describes the detail features of jammer more accurately.
\item The proposed algorithm is relying only on the locally observed information and does not need to estimate the  jamming patterns and parameters the jammer in advance, i.e., it is model-free, which can be widely used in various anti-jamming scenarios.
\end{itemize}

Note that the most related work is \cite{bb_DQN1}, which also adopted deep reinforcement learning to investigate the anti-jamming problems. The main differences in this work are as follows: i) the environment state is presented by extracting features of signal-to-interference-plus-noise ratio (SINR) and primary user occupancy in \cite{bb_DQN1}, while it is presented by the raw spectrum information in this work, and ii) it requires the jammer to have the same channel-slot transmission structure with the users in \cite{bb_DQN1}. On the contrary, this requirement does not hold in our work, which makes the proposed approach more general.

\vspace{-0.2cm}
\section{System Model and Problem Formulation}

\begin{figure}[!t]
\centering
\includegraphics[width=0.8\linewidth]{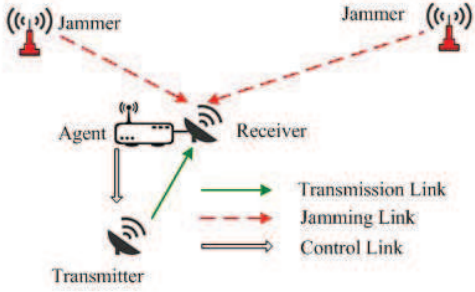}
\caption{System model.}
\label{systemmodel}
\end{figure}

We consider the transmission of one user (a transmitter-receiver pair) against one or several jammers, as shown in Fig.~\ref{systemmodel}. The agent, which is disposed at the receiving end, sends anti-jamming strategies to the transmitter through a reliable control link. Jammers may adopt fixed, random, or possibly intelligent jamming patterns. However, we do not analyze the specific jamming models, but obtain the optimal anti-jamming strategies based on the raw spectrum information.

While the receiver receives the desired signal, the agent continuously senses the whole communication bands and stores the sensed values. Denote the spectrum vector as ${\bf {P}}_t = \left\{ p_{t,1}, \ p_{t,2},\ \cdots ,\ p_{t,N} \right\}$, where ${p_{t,n}}$ is the power of frequency ${n}$ at time ${t}$ and $N$ is the number of sampling points in frequency space. In order to sufficiently use history spectrum information, a two-dimensional matrix, which describes time-frequency features of spectrum environment, is expressed as:

\begin{equation}
\label{eq_s_def}
{\bf {S}}_t=
	\begin{bmatrix}
	{\bf {P}}_{t-1}\\
    {\bf {P}}_{t-2}\\
   		\par\vdots\\
    {\bf {P}}_{t-M}\\
	\end{bmatrix}
=
\begin{bmatrix}
	p_{t-1,1}&p_{t-1,2}&\cdots&p_{t-1,N}\\
    p_{t-2,1}&p_{t-2,2}&\cdots&p_{t-2,N}\\
   	 \vdots   & \vdots & \ddots  & \vdots  \\
    p_{t-M,1}&p_{t-M,2}&\cdots&p_{t-M,N}\\
\end{bmatrix}.
\end{equation}
It is noted that  ${{\bf {S}}_t}$ contains all the spectrum information until time $t$, as ${M}$ tends to infinity. However, the difficulty of the decision optimization problem is significantly increased with the increase of ${M}$. Therefore, ${M}$ can take an appropriate value, which would be determined by the time-varying characteristics of the spectrum environment.

\begin{figure}[!t]
\centering
\includegraphics[width=0.9\linewidth]{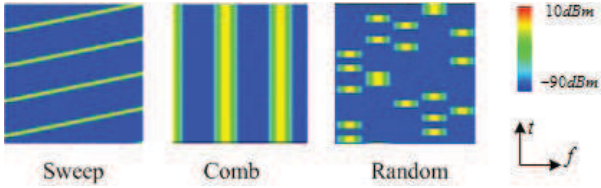}
\caption{Thermodynamic chart of various jamming pattern.}
\label{pattern}
\end{figure}

To illustrate the rationality of using ${{\bf {S}}_t}$ as the basis of anti-jamming decision-making, we give the thermodynamic charts of the ${{\bf {S}}_t}$ matrix of several common jamming patterns, also known as spectrum waterfall  \cite{bb_waterfall}, as shown in Fig.~\ref{pattern}. Taking the swept jamming as example, we can accurately determine the frequency range and intensity (color) of jamming at the next moment by looking at the thermal chart, which also means we can determine the anti-jamming strategy accordingly.

In the unknown and dynamic environment, we do not consider estimation-based anti-jamming strategies. Instead,  define ${{\bf {S}}_t}$ as the environment state, and then consider a dynamic decision problem in which the agent (anti-jamming user) interacts with an environment through a sequence of observations of environment (${{\bf {S}}_t}$), actions ($a_t$) and rewards  ($r_t$). Specifically, an action $a$ can be a combination decisions of frequency, power, coding schemes, spread spectrum, and other kinds of anti-jamming decisions, e.g., $a=(f,p)$ represents the combination actions of frequency ($f$) and power ($p$). The rewards associated with the actions and environment is defined as:
\begin{equation}
\label{eq_rewards}
r(a,{\bf {S}})= \left\{
\begin{array}{lr}
R(a)-\lambda \delta & \beta(a,{\bf {S}})\ge \beta_{th}(a)\\
0  & \beta(a,{\bf {S}})<\beta_{th}(a)
\end{array}
\right.,
\end{equation}
where $R(a)$ is the bit rate when the action $a$ is selected, $\lambda$ is the cost when action changes, $\delta$ is an indication of action change ($\delta=1$ if $a_t \neq a_{t-1}$; $\delta=0$ if $a_t = a_{t-1})$,  $\beta(a,{\bf {S}})$ is the received signal to interference plus noise ratio (SINR) in state ${\bf {S}}$ with action $a$. $\beta_{th}(a)$ is the required SINR threshold for successful transmission. Note that $R(a)$ and $\beta_{th}(a)$ are modeled as a function of action $a$, the reason is as follows: the bit rate and SINR requirements change  for different anti-jamming strategies, such as forward error correction and spread spectrum schemes.

 Then, the goal of the agent is to select anti-jamming actions in a fashion that maximizes cumulative future reward $R_t= \sum_{i=0}^{\infty} {\gamma^ir_{t+i+1}}$, where $\gamma$ is the discount factor.  One way of achieving this goal is to compute the following optimal action-value (also known as $Q$) function \cite{bb_Qlearning}:
\begin{equation}
\label{eq_qf_tradition}
Q^{\ast}( {\bf {S}} ,a)=\mathop {\max }\limits_{\pi}{E \left\{
R_t
 \mid {\bf {S}}_t={\bf {S}}, a_t=a, \pi \right\} },
\end{equation}
where the anti-jamming policy $\pi=P(a\,|\,{\bf {S}})$ refers to a probability distribution over the actions.  Based on the Bellman equation,
\begin{equation}
\label{eq_qf_update}
Q^{\ast}( {\bf {S}} ,a)=E \left\{
r+\gamma \mathop{\max}\limits_{a'}{Q^{\ast}({\bf {S}}',a') \mid  {\bf {S}}, a}
\right\}.
\end{equation}


\section{Anti-jamming Communication Scheme}
\begin{figure*}[htbp]
\centering
\includegraphics[width=0.7\linewidth]{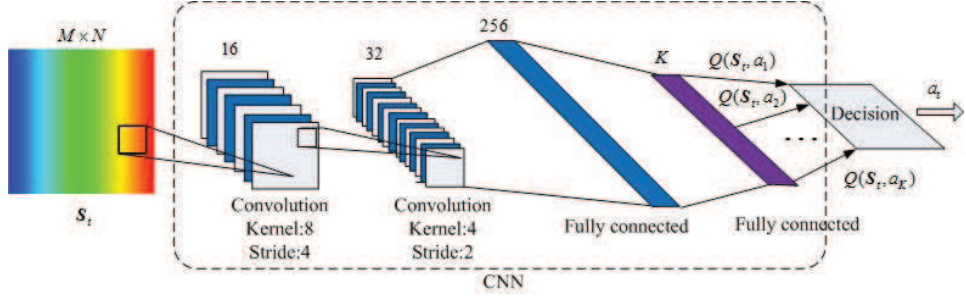}
\caption{Decision network of the DAQN.}
\label{decision}
\end{figure*}

\begin{figure}[!t]
\centering
\includegraphics[width=0.7\linewidth]{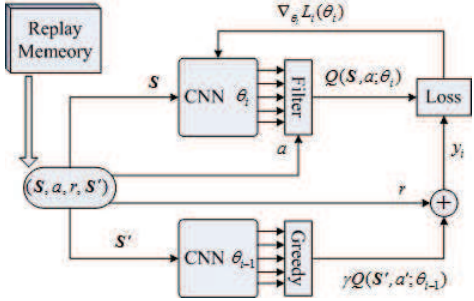}
\caption{Updating network of the DAQN.}
\label{update}
\end{figure}

The traditional Q-learning is unable to cope with the anti-jamming problem described in section \uppercase\expandafter{\romannumeral2}, as the state space size of ${{\bf {S}}_t}$ is almost infinite. In order to solve this problem, a deep anti-jamming Q-network (DAQN) is constructed to address the interactive decision-making problem with raw spectrum information input, which contains decision network and update network, as shown in Fig.~\ref{decision} and Fig.~\ref{update} respectively. We use a deep convolutional neural network (CNN) to approximate the optimal action-value as shown in decision network, where the input state ${\bf {S}}_t$ is represented by a thermal chart of $M \times N$ pixels. After the processing of two  convolutional layers and two fully connected layers, the output is the estimated $Q$ function, where $K$ is the size of action space. At last, the decision layer outputs the corresponding action based on the estimated $Q$ function.

However,  reinforcement learning is known to be unstable or even to diverge when a nonlinear function approximator \cite{bb_DQN}, such as the neural network is used to represent the $Q$ function. The main reason is correlation during the learning process. The idea of experience replay is adopted to address these instabilities as shown in the update network. To perform experience replay, we store the agent's experiences $e_t=({\bf {S}}_t,a_t,r_t,{\bf {S}}_{t+1})$ at each time-step $t$ in data set $D_t=(e_1,\cdots\,e_t)$. When the experience pool is big enough, we construct target values $r+\gamma \mathop {\max }\limits_{a'} {Q({\bf {S}}',a')}$ by randomly choosing elements in a uniform distribution  $({\bf {S}},a,r,{\bf {S}}') \sim U(D)$, which reduces the correlation during sequential observation. The Q-learning update at iteration $i$ uses the following loss function:

\begin{equation}
\label{eq_q_update}
L_i(\theta_i)=E_{({\bf {S}},a,r,{\bf {S}}') \sim U(D)}
\left[
\left(y_i-Q({\bf {S}},a;\theta_i)\right)^2
\right],
\end{equation}
where $\theta_i$ is the parameter of Q-network at iteration $i$ and $y_i=r+\gamma \mathop {\max }\limits_{a'} {Q({\bf {S}}',a';\theta_{i-1})}$ is target value computed by Q-network parameter $\theta_{i-1}$ with greedy strategy. By assuming that $y_i$ is the expected output of CNN with network weight $\theta_i$ when the input is ${\bf {S}}$, we calculate the difference between real output $Q({\bf {S}},a;\theta_i)$ and target value $y_i$ to determine the update of network parameters. Differentiating the loss function with respect to the weights, we arrive at the following gradient:

\begin{equation}
\label{eq_gradient}
\nabla_{\theta_i} L_i(\theta_i)=
E_{({\bf {S}},a,r,{\bf {S}}')}
\left[
\left(y_i-Q({\bf {S}},a;\theta_i)\right)
\nabla_{\theta_i}Q({\bf {S}},a;\theta_i)
\right].
\end{equation}
According to the gradient descent algorithm, the network weight $\theta_i$ is updated according to (\ref{eq_gradient}). Although there are two CNN networks with different weights, as shown in Fig.~\ref{update}, the actual implementation requires only one CNN network, as the computing of target values and the updating of network weights are in different stages. The algorithm for anti-jamming communication based on deep reinforcement learning is presented in Algorithm 1.

\begin{figure}[tb]
\rule{\linewidth}{1pt}
\emph{\textbf{Algorithm 1}: Deep Anti-jamming Reinforcement Learning  Algorithm (DARLA)}\\
\rule{\linewidth}{1pt}
\begin{algorithmic}
\STATE \textbf{Initialize :} Set $D=\emptyset$, $\epsilon=1$, Set $\theta$ with random weights, Sense initial environment ${\bf {S}}_1$.
\STATE \textbf{For} $t=1,T$ \textbf{do}
\STATE \qquad With probability $\epsilon$, select a random action $a_t$
\STATE \qquad otherwise, select $a_t=\mathop {\arg\max }\limits_{a} {Q({\bf {S}}_t,a;\theta)}$
\STATE \qquad Execute action $a_t$ and compute $r_t$ and observe  ${\bf {S}}_{t+1}$
\STATE \qquad Store transitions $({\bf {S}}_t,a,r,{\bf {S}}_{t+1})$  in $D$
\STATE \qquad \textbf{If} $Sizeof(D)>\mathcal{N}$(Enough amount of transitions)
\STATE \qquad \qquad Sample random minibatch of transistions
\STATE \qquad \qquad $({\bf {S}},a,r,{\bf {S}}')$ from $D$
\STATE \qquad \qquad Compute $y_i=r+\gamma \mathop {\max }\limits_{a'} {Q({\bf {S}}',a';\theta)}$
\STATE \qquad \qquad Compute gradient based on Eq.(\ref{eq_gradient}) and Update $\theta$
\STATE \qquad \textbf{End If}
\STATE \qquad Calculate $\epsilon=\text{max}(0.1, \epsilon-\Delta\epsilon)$
\STATE \textbf{End For}
\end{algorithmic}
\rule{\linewidth}{1pt}
\end{figure}

\vspace{-0.3cm}
\section{Numerical Results and Discussions}



In the simulation setting, the user and the jammer combat with each other in a frequency band of 20MHz, where the frequency resolution of spectrum sensing is 100kHz. The user performs a full band sensing every 1ms and retains the spectrum data within the 200ms. Hence, the size of matrix ${\bf {S}}_t$ is $200 \times 200$. The bandwidth of user signal is 4MHz, and the center frequency is allowed to change in each 10ms with the step of 2MHz, which means $K=9$. Both signal and jamming are raised cosine waveform with roll-off factor $\alpha=0.3$, in which jamming power is 30dBm and signal power is 0dBm. The demodulation threshold $\beta_{th}$ at all frequency is set to be 10dB, and the cost of action change $\lambda$ is set to be $0.2R(a)$.

Four kinds of jamming patterns are given for simulation: \expandafter{\romannumeral1}) Sweep jamming (sweep speed is 1GHz/s); \expandafter{\romannumeral2}) Comb jamming (three fixed frequency signals at 2MHz, 10MHz, and 18MHz); \expandafter{\romannumeral3}) Random jamming (frequency is randomly changed every 20ms with the step of 4MHz); \expandafter{\romannumeral4}) Intelligent jamming (the jammer continuously observes the probability that the user signal appears at each frequency point, and chooses the largest one as jamming channel). For all Jamming patterns, the instantaneous bandwidth of the jamming is set to be 4MHz.

\begin{figure}[!t]
\centering
\includegraphics[width=0.7\linewidth]{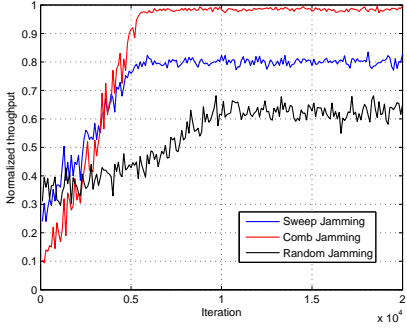}
\caption{Normalized throughput under different jamming patterns.}
\label{throughput}
\end{figure}

\begin{figure}[!t]
\centering
\includegraphics[width=0.7\linewidth]{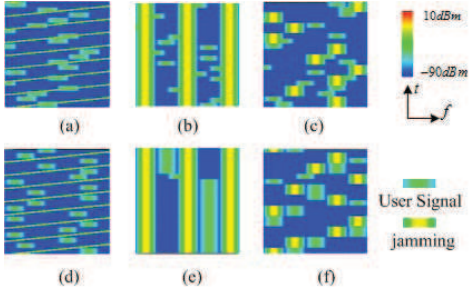}
\caption{Environmental states at initial and convergent stages under different jamming patterns.}
\label{result}
\end{figure}

\begin{figure}[!t]
\centering
\includegraphics[width=0.75\linewidth]{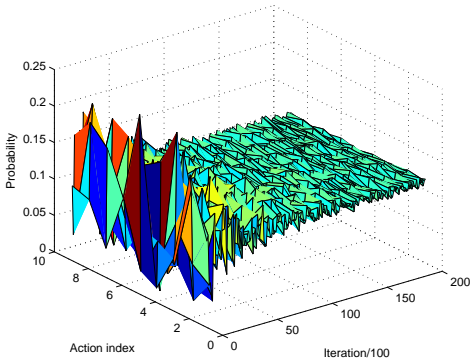}
\caption{Probability of user actions during learning.}
\label{prob}
\end{figure}

The normalized average throughput of legitimate user under different jamming patterns is given in Fig.~\ref{throughput}.  It is shown that the anti-jamming ability of user has been improved significantly with the proposed DARLA learning. Especially in the case of comb jamming, the normalized throughput is close to one after convergence, which indicates that the jamming is almost completely avoided.

Environmental states at initial stages under sweep, comb and random jamming patterns are given in Fig.~\ref{result}(a), (b), and (c) respectively, and the converging states are given in Fig.~\ref{result}(d), (e), and (f) respectively. These states that contain time-frequency information can clearly reflect the past actions of user and jamming. Taking sweep jamming as an example, at the beginning of the learning procedure, the user adopts randomized action as it is unfamiliar with environment (the locations of rectangular blocks are randomly distributed), and after convergence, the frequency is properly changed before the jamming arrives (the rectangular blocks are distributed according to the slashes).

With regard to the intelligent jamming, since the probability distribution of user actions is the basis for jammer to release jamming, the best strategy for user is that the probability of each action is almost identical. The simulation results in Fig.~\ref{prob} show the probabilities of each action being selected during the learning procedure, which is consistent with our analysis.

\vspace{-0.2cm}
\section{Conclusion}
In this letter, we investigated the anti-jamming problem in unknown and dynamic environment. Aiming at employing the waterfall spectrum information directly,  we constructed a deep anti-jamming Q-network  to handle the complex interactive decision-making problem with infinite number of states. Then, a deep anti-jamming  reinforcement learning  algorithm was proposed. Using the proposed learning algorithm, the user is able to learn the best anti-jamming strategy by constantly trying various actions and sensing the spectrum environment. Simulation results in various scenarios are presented to validate the proposed anti-jamming communication approach. Future work on designing multi-user deep anti-jamming reinforcement learning algorithms  is ongoing.

\ifCLASSOPTIONcaptionsoff
  \newpage
\fi

\end{document}